# Generation of sub-20-fs pulses from a graphene mode-locked laser

**FERDA CANBAZ,[1] NURBEK KAKENOV,[2] COSKUN KOCABAS,[2] UMIT DEMIRBAS,[3] AND ALPHAN SENNAROGLU[1,4*]**

[1]*Laser Research Laboratory, Departments of Physics and Electrical-Electronics Engineering, Koç University, İstanbul 34450, Turkey*
[2]*Department of Physics, Bilkent University, Ankara 06800, Turkey*
[3]*Department of Electrical and Electronics Engineering, Antalya International University, Antalya 07190, Turkey*
[4]*Koç University Surface Science and Technology Center (KUYTAM), Rumelifeneri, Sarıyer, Istanbul 34450, Turkey*
*[*asennar@ku.edu.tr](mailto:asennar@ku.edu.tr)*

**Abstract:** We demonstrate, what is to our knowledge, the shortest pulses directly generated to date from a solid-state laser, mode locked with a graphene saturable absorber (GSA). In the experiments, a low-threshold diode-pumped $Cr^{3+}$:LiSAF laser was used near 850 nm. At a pump power of 275 mW provided by two pump diodes, the $Cr^{3+}$:LiSAF laser produced nearly transform-limited, 19-fs pulses with an average output power of 8.5 mW. The repetition rate was around 107 MHz, corresponding to a pulse energy and peak power of 79 pJ and 4.2 kW, respectively. Once mode locking was initiated with the GSA, stable, uninterrupted femtosecond pulse generation could be obtained. In addition, the femtosecond output of the laser could be tuned from 836 nm to 897 nm with pulse durations in the range of 80-190 fs. We further performed detailed mode locking initiation tests across the full cavity stability range of the laser to verify that pulse generation was indeed started by the GSA and not by Kerr lens mode locking.

## References and links

1. P. Avouris, "Graphene: Electronic and Photonic Properties and Devices," Nano Lett. **10**, 4285-4294 (2010).
2. K. S. Novoselov, V. I. Falko, L. Colombo, P. R. Gellert, M. G. Schwab, and K. Kim, "A roadmap for graphene," Nature **490**, 192-200 (2012).
3. J. M. Dawlaty, S. Shivaraman, M. Chandrashekhar, F. Rana, and M. G. Spencer, "Measurement of ultrafast carrier dynamics in epitaxial graphene," Appl. Phys. Lett. **92**, 042116 (2008).
4. A. B. Kuzmenko, E. van Heumen, F. Carbone, and D. van der Marel, "Universal Optical Conductance of Graphite," Phys. Rev. Lett. **100**, 117401 (2008).
5. I. H. Baek, H. W. Lee, S. Bae, B. H. Hong, Y. H. Ahn, D.-I. Yeom, and F. Rotermund, "Efficient Mode-Locking of Sub-70-fs Ti:Sapphire Laser by Graphene Saturable Absorber," Appl. Phys. Express **5**, 032701 (2012).
6. M. N. Cizmeciyan, J. W. Kim, S. Bae, B. H. Hong, F. Rotermund, and A. Sennaroglu, "Graphene mode-locked femtosecond Cr:ZnSe laser at 2500nm," Opt. Lett. **38**, 341-343 (2013).
7. J. Ma, H. T. Huang, K. J. Ning, X. D. Xu, G. Q. Xie, L. J. Qian, K. P. Loh, and D. Y. Tang, "Generation of 30 fs pulses from a diode-pumped graphene mode-locked Yb:CaYAlO4 laser," Opt. Lett. **41**, 890-893 (2016).
8. N. Tolstik, E. Sorokin, and I. T. Sorokina, "Graphene mode-locked Cr:ZnS laser with 41 fs pulse duration," Opt. Express **22**, 5564-5571 (2014).
9. U. Demirbas and I. Baali, "Power and efficiency scaling of diode pumped Cr:LiSAF lasers: 770-1110 nm tuning range and frequency doubling to 387-463 nm," Opt. Lett. **40**, 4615-4618 (2015).
10. F. Druon, F. Balembois, and P. Georges, "New laser crystals for the generation of ultrashort pulses," C. R. Phys. **8**, 153-164 (2007).
11. S. Uemura and K. Torizuka, "Generation of 10 fs pulses from a diode-pumped Kerr-lens mode-locked Cr : LiSAF laser," Jpn. J. Appl. Phys. 1 **39**, 3472-3473 (2000).
12. F. Canbaz, N. Kakenov, C. Kocabas, U. Demirbas, and A. Sennaroglu, "Graphene mode-locked Cr:LiSAF laser at 850 nm," Opt. Lett. **40**, 4110-4113 (2015).

13. O. Salihoglu, S. Balci, and C. Kocabas, "Plasmon-polaritons on graphene-metal surface and their use in biosensors," Appl. Phys. Lett. **100**, 213110 (2012).
14. H. A. Haus, "Mode-locking of lasers," IEEE J. Sel. Top. Quant. **6**, 1173-1185 (2000).
15. I. T. Sorokina and K. L. Vodopyanov, *Solid-State Mid-Infrared Laser Sources*, Topics in Applied Optics (Springer), Vol. 89.

## 1. Introduction

The unique electrical and optical characteristics of graphene have enabled its widespread use in different applications [1, 2]. In particular, due to its picosecond recovery time and strong absorption saturation, graphene has been successfully demonstrated as a fast saturable absorber for the generation of ultrashort optical pulses from lasers [3]. The gapless band structure of this important nanostructured material further leads to a nearly constant absorption band (about 2.3% per pass) [4], extending over a broad wavelength range (0.7-25 μm). This enables the use of graphene saturable absorbers (GSAs) in lasers operating at different wavelengths. Hence, provided that the graphene layer is transferred onto a transparent substrate, the same GSA can be used to mode lock lasers over a very broad wavelength range, unlike semiconductor saturable absorber mirrors (SESAMs) or single-walled carbon nanotubes (SWCNTs), which rely on resonant absorption over a more limited wavelength range. To date, GSAs have been used to initiate mode-locked operation of lasers operating between 800 nm and 2500 nm [5, 6].

The ultrabroad saturable absorption band of GSAs should in principle impose no bandwidth limitation during mode locking, leading to pulse durations only limited by the amplification bandwidth of the gain medium and dispersion of the cavity. To date, the shortest pulses obtained from a GSA mode-locked laser had a duration of 30 fs ($Yb^{3+}$:$CaYAlO_4$ laser) [7]. In another study, 41-fs pulses were generated from a GSA mode-locked $Cr^{2+}$:ZnS laser at 2400 nm, corresponding to 5.1 optical cycles [8]. In our current GSA mode locking experiments, we used a $Cr^{3+}$:LiSAF gain medium to demonstrate ultrashort pulse generation, since the amplification bandwidth of $Cr^{3+}$:LiSAF is broad enough to support sub-10-fs pulses [9]. Furthermore, in comparison with $Ti^{3+}$:sapphire lasers $Cr^{3+}$:LiSAF has a larger stimulated emission-lifetime product, making it possible to achieve lower lasing thresholds [10]. As a result, $Cr^{3+}$:LiSAF gain medium can be pumped with low-cost, widely available, low-power red diode lasers to obtain efficient power performance. This attractive feature makes it possible to develop robust, low-cost femtosecond sources in the near infrared. In an earlier study, pulses as short as 10 fs were generated from a Kerr-lens mode-locked $Cr^{3+}$:LiSAF laser [11]. In our previous study, we demonstrated GSA-initiated mode locking of a $Cr^{3+}$:LiSAF laser for the first time and generated 68-fs pulses at 850 nm [12].

In this Paper, we report on the direct generation of sub-20-fs pulses from a GSA mode-locked $Cr^{3+}$:LiSAF laser. The total group delay dispersion (GDD) inside the cavity was controlled by using chirped mirrors and a prism pair. Once the focusing on the GSA was optimized, mode-locked operation could be readily obtained by translating the output coupler. At the input pump power of 275 mW, we obtained 19-fs pulses with an average power of 8.5 mW. In addition, the femtosecond output of the laser could be tuned from 836 nm to 897 nm with pulse durations in the range of 80-190 fs. Furthermore, we performed detailed mode locking initiation tests across the full cavity stability range of the laser to verify that pulse generation was indeed started by GSA and not by Kerr lens mode locking (KLM). To the best of our knowledge, these represent the shortest pulses generated to date with a GSA mode-locked solid-state laser.

## 2. Experimental details and results

A schematic of the sub-20-fs, GSA mode-locked $Cr^{3+}$:LiSAF laser is shown in Fig. 1. The Brewster cut gain crystal with a length of 6 mm and chromium concentration of 1.5% was end pumped with two polarization-coupled, single-mode diode lasers (D1 and D2 in Fig. 1) operating around 660 nm. The total absorption of the gain crystal was 98% at 660 nm. The output beam of each diode laser was collimated with an aspheric lens (L1 and L2 in Fig. 1, focal length=4.5 mm). The diode output beams were then combined with a polarizing beam splitter (PBS), giving a total input pump power of 275 mW. The combined pump beam was focused inside the $Cr^{3+}$:LiSAF gain crystal with a 60-mm focal-length lens. The laser resonator consisted of two concave focusing mirrors, each with a radius of curvature (ROC) of 75 mm (M1 and M2), a pair of fused silica prisms (FSP1 and FSP2), a highly reflecting end mirror (M5), and an output coupler with 0.5% transmission (OC). A second intra-cavity beam waist was formed by using two additional curved high reflectors (M3 and M4, ROC=75 mm). The monolayer GSA was placed between M3 and M4 to increase the beam fluence on the saturable absorber. The GSA used in our experiments was prepared by chemical vapor deposition as described previously [13] and was transferred onto an infrasil substrate. As described in Ref [12], the saturation fluence and recovery time were 28 $\mu J/cm^2$ and 1.6 ps, respectively. Furthermore, previous Raman measurements indicated that a monolayer graphene was deposited on the infrasil substrate [12]. To reduce reflection losses, the GSA was further placed at Brewster incidence. The prism pair (FSP1 and FSP2) with a tip-to-tip separation of 29 cm was placed in the high reflector (HR) arm for dispersion control. HR and OC arm lengths were 88 cm and 41 cm, respectively, whereas the total cavity length was 138 cm (pulse repetition rate=107 MHz).

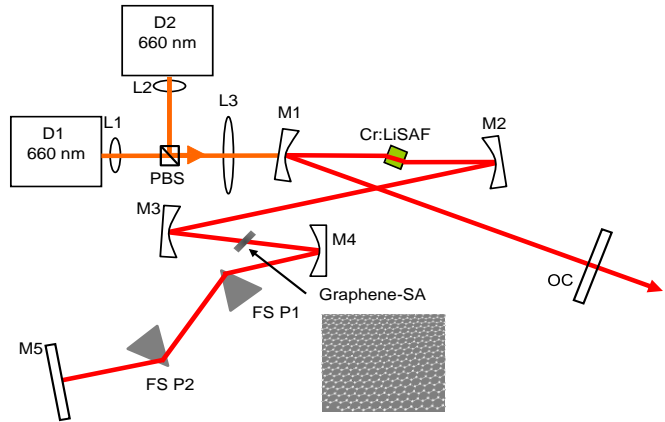

Fig. 1 Schematic of the diode-pumped, GSA mode-locked $Cr^{3+}$:LiSAF laser.

Figure 2(a) shows the power efficiency curves for the continuous-wave (cw) $Cr^{3+}$:LiSAF laser (containing the prism pair and an infrasil substrate) with different output couplers. In the mode locking experiments, in order to maintain a high intracavity power and to remain near the optimum output coupling of the resonator, we used the 0.5 % output coupler. By using the 0.5% output coupler, as high as 75 mW of output power was obtained near 850 nm with 275 mW of incident pump power [Fig. 2(a)]. After the insertion of the GSA, the output power of the laser decreased to 6.2 mW (0.5% output coupler), as shown in Fig. 2(b). By using the power efficiency data [Figs. 2(a) and 2(b)] as well as the pump threshold data, we estimated the round-trip loss of the free-running laser and GSA to be 0.4% and 4.1%, respectively.

After the insertion of the GSA into the resonator, femtosecond pulse generation could be readily initiated by translating the output coupler. Pulse characterization was also performed

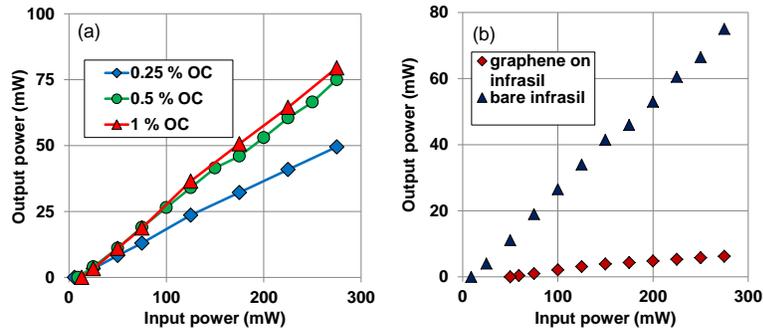

Fig. 2 (a) Continuous-wave (cw) power efficiency curves of the $Cr^{3+}$:LiSAF laser with different output couplers. (b) Cw efficiency data of the laser containing a bare infrasil substrate and infrasil with graphene. The output coupler has 0.5% transmission. The cavity also contains the prism pair and infrasil substrate in all cases.

at the maximum pump power by using the 0.5% output coupler. Figure 3(a) shows the optical spectrum of the shortest pulses we generated and the corresponding estimated group delay dispersion (GDD) as a function of wavelength. Once mode locking was initiated, the output power increased from 6.2 mW to 8.5 mW due to the saturation of the GSA. Based on the measured optical spectrum, we estimated the shortest transform-limited pulse duration and the corresponding time-bandwidth product to be 17 fs and 0.41, respectively.

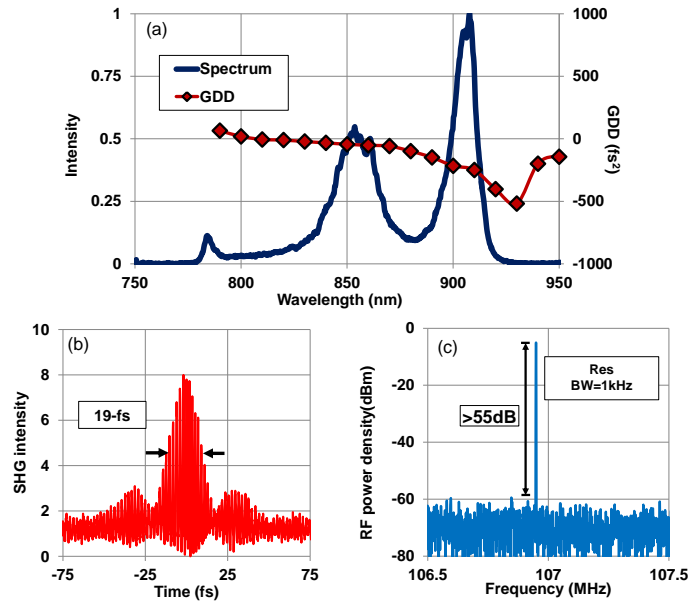

Fig. 3 (a) Optical spectrum and estimated group delay dispersion (GDD), (b) interferometric autocorrelation, and (c) radio frequency spectrum of the generated pulses with GSA-mode-locked $Cr^{3+}$:LiSAF laser at the input pump power of 275 mW.

The experimentally measured interferometric autocorrelation trace [Fig. 3(b)] yielded a pulse duration of 19 fs with a measured time bandwidth product of 0.46, suggesting that the generated pulses were nearly transform limited. Furthermore, in the RF spectrum measurements, the sidebands were at least 55 dB below the fundamental tone at the repetition rate of 107 MHz as can be seen from Fig. 3(c). In mode locking experiments, to balance the Kerr nonlinearities and to generate solitary pulses, dispersion management was employed by using dispersion control mirrors (DCM, M1 and M2, each with −80±10 $fs^2$ of GDD per bounce and M5 with −40±10 fs$^2$ of GDD per bounce) and a pair of fused silica prisms (round trip GDD~-500fs$^2$). The material GDD contributions of the gain crystal, fused silica prism material, and the infrasil substrate were +24, +40, and +40 fs$^2$, respectively. The net average round-trip GDD over the bandwidth was hence estimated as -50 fs$^2$. We further estimated the net round-trip GDD from the soliton area theorem [14], by using the pulse duration, intracavity pulse energy and the previously reported nonlinear refractive index of the Cr$^{3+}$:LiSAF gain medium (0.8x10$^{-16}$ cm$^2$/W [15]). This gave a round-trip GDD of -20fs$^2$. The discrepancy may be due to the fact that the actual GDD of the DCMs varies somewhat depending on the angle of incidence.

Broad tunability of the mode-locked laser could be very useful in many applications such as multi-photon microscopy or pump-probe spectroscopy. In the experiments, we introduced an adjustable slit into the GSA-mode-locked laser between the second prism P2 and the end high reflector (M5) to narrow the spectral bandwidth of the pulses and to investigate the tuning capability. The central wavelength of the mode-locked spectrum was adjusted by changing the position of the adjustable slit. As can be seen from the spectra displayed in Fig. 5, continuous femtosecond tuning could be achieved at wavelengths from 836 nm to 897 nm. At a fixed slit width that was adjusted to give about 75-fs pulses near 875 nm, the pulse duration remained in the 75-190 fs range between 836 and 892 nm with time-bandwidth products between 0.53 and 0.77. The pulses became broader near the long wavelength edge due to the steep GDD of the cavity optics [see Fig. 3(a)]. The tuning data of Fig. 4 were recorded with 275 mW of input pump power.

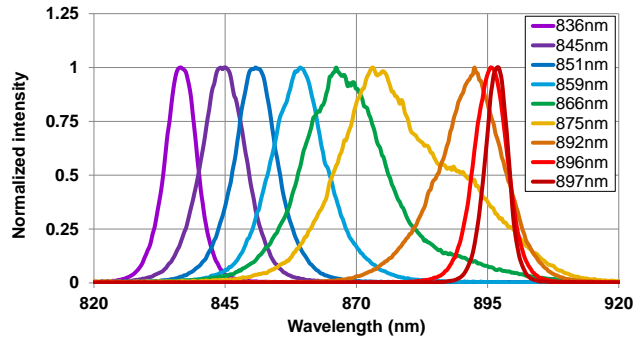

Fig. 4 Femtosecond tuning data of the GSA-mode-locked Cr$^{3+}$:LiSAF laser.

To verify that mode locking was indeed initiated by the fast saturable absorber action in graphene, we checked the initiation mechanism of mode locking across the full cavity stability range. In particular, we varied the position of the curved mirror M2 at 10-µm intervals across the full stability range and at each point, we checked whether or not mode locking could be initiated by translating the output coupler. In the control experiment, the infrasil substrate with GSA was replaced with a bare infrasil substrate and the same test was repeated. Figures 5(a) and (b) show the measured output power as a function of the M2 position for bare infrasil substrate and infrasil substrate with GSA. Figures 5(c) and 5(d) show at which points within the stability range mode locking could be initiated. The values of 1, 0.5 and 0 indicate M2 positions with stable mode locking, only flashing of mode-locked operation, and no mode locking, respectively. Note that with the bare infrasil substrate, mode locking could be initiated at far fewer points [see Fig. 5(c)]. In this case, the initiation mechanism was Kerr lens mode locking (KLM) that requires critical cavity alignment. With the GSA on infrasil [Fig. 5(d)], we were able to obtain stable mode-locked operation over the full stability range except at the edges, indicating that the initiation mechanism had no sensitivity to critical cavity alignment, and clearly verifying that mode locking was initiated by the fast saturable absorber action in the GSA.

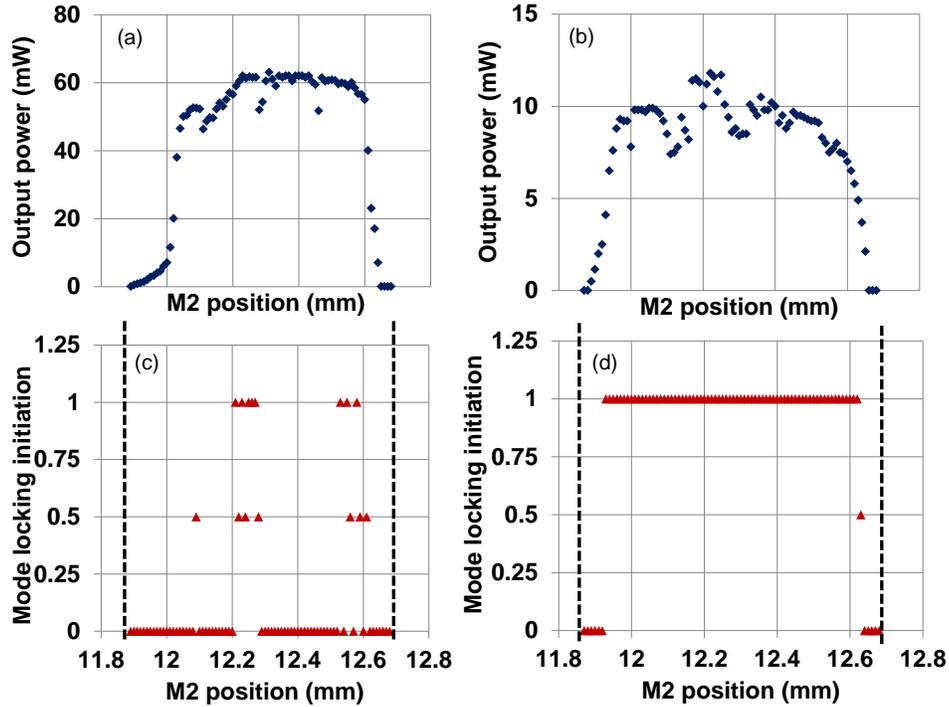

Fig. 5 Measured output power of the $Cr^{3+}$:LiSAF laser as a function of the M2 position with (a) a bare infrasil substrate and (b) an infrasil substrate with graphene. Mode locking initiation maps of the resonator with (c) a bare infrasil substrate and (d) an infrasil substrate with graphene.

Also note that, the shortest pulse data displayed in Fig. 3 were recorded at the M2 location (12.48 mm) where only the GSA took part in the initiation of femtosecond pulse train without any KLM effect. Also note that as the output power data of Fig. 5(b) were measured, transients caused by the moving output coupler led to damage on the GSA at several M2 positions. The position of the GSA was then slightly changed, leading to the power fluctuation seen in Fig. 5(b). However, at the M2 position where the shortest pulse data were recorded, no damage was observed on the GSA up to the maximum fluence level of 1264 µJ/cm², estimated by assuming a mode-locked output power of 8.5 mW and by using the other relevant parameters of the resonator.

In conclusion, we have generated, what is to our knowledge, the shortest pulses, directly from a graphene mode-locked solid-state laser. The graphene mode-locked $Cr^{3+}$:LiSAF gain medium employed in the experiments, was operated in a low-threshold resonator configuration to generate stable, 19-fs pulses near 850 nm with only 275 mW of pump power. In addition, the femtosecond output of the laser could be tuned from 836 nm to 897 nm. Finally, we used a definitive experimental test to clearly identify the initiation mechanism for mode locking and experimentally demonstrated that the femtosecond pulse train was initiated solely by the saturable absorber action in graphene.

### 3. Funding

Coskun Kocabas acknowledges the support of the European Research Council (ERC) Consolidator Grant ERC-682723 SmartGraphene.